\documentclass[superscriptaddress,reprint,amsmath,amssymb,aps]{revtex4-2}
\usepackage{graphicx}% Include figure files
\usepackage{dcolumn} % Align table columns on decimal point
\usepackage{bm}      % bold math
\usepackage{hyperref}% add hypertext capabilities
\usepackage{mydef}
\usepackage{xcolor}

\usepackage{float}
\usepackage{stmaryrd}
\begin{document}
\preprint{APS/123-QED}
\title{Geometric Effects in Large Scale Intracellular Flows}

\author{Olenka Jain}
\affiliation{Lewis- Sigler  Institute  for  Integrative  Genomics,  Princeton  University, Princeton, NJ 08544}
\affiliation{Center for Computational Biology, Flatiron Institute, New York, 10010, USA}
\author{Brato Chakrabarti}
\affiliation{International Center for Theoretical Sciences, Bengaluru, 560089, India}%
\author{Reza Farhadifar}
\affiliation{Center for Computational Biology, Flatiron Institute, New York, 10010, USA}
\author{Elizabeth R. Gavis}
\affiliation{Department of Molecular Biology, Princeton University, Princeton, NJ 08544}
\author{Michael J. Shelley}
\affiliation{Center for Computational Biology, Flatiron Institute, New York, 10010, USA}
\affiliation{Courant Institute, New York University, New York, NY 10012, USA}
\author{Stanislav Y. Shvartsman}
\affiliation{Center for Computational Biology, Flatiron Institute, New York, 10010, USA}
\affiliation{Department of Molecular Biology, Princeton University, Princeton, NJ 08544}
\affiliation{Lewis- Sigler  Institute  for  Integrative  Genomics,  Princeton  University, Princeton, NJ 08544}

\date{\today}

\begin{abstract}
This work probes the role of cell geometry in orienting self-organized fluid flows in the late stage \textit{Drosophila} oocyte. Recent theoretical work has shown that a model, which relies only on hydrodynamic interactions of flexible, cortically anchored microtubules (MTs) and the mechanical loads from molecular motors moving upon them, is sufficient to generate observed flows. While the emergence of flows has been studied in spheres, oocytes change shape during streaming and it was unclear how robust these flows are to the geometry of the cell. 
Here we use biophysical theory and computational analysis to investigate the role of geometry and find that the axis of rotation is set by the shape of the domain and that the flow is robust to biologically relevant perturbations of the domain shape. Using live imaging and 3D flow reconstruction, we test the predictions of the theory/simulation, finding consistency between the model and live experiments, further demonstrating a geometric dependence on flow direction in late-stage \textit{Drosophila} oocytes.
\end{abstract}

\maketitle
\section{\label{sec:level1} Introduction}
Symmetry breaking is a fundamental process in biology, critical for development and organization, from subcellular structures to whole body organization. Biological systems employ myriad internal and external cues, such as chemical gradients and force imbalances to break symmetries. Here we examine the role of a less well characterized source of asymmetry: the shape of the domain in which dynamics play out \cite{Bialecki2017,Pintado2017}. Our work is motivated by self-organized cell-scale fluid flows, referred to as cytoplasmic streaming, associated with a cortically bound bed of flexible microtubule fibers under motor protein loads in the \textit{Drosophila} oocyte. The flows are critical for transporting material in the late-stage \textit{Drosophila} oocyte, where the large size of the cell limits the efficiency of diffusion based transport \cite{Quinlan2016,Dutta2024,Forrest2003}. Without flows, specific mRNA molecules that are critical for setting up the chemical gradients that prepattern the \textit{Drosophila} embryo, are improperly localized and result in disrupted development. Understanding how such self-organizing systems pick a direction for flow is critical for quantifying their transport and mixing properties. Here, we examine whether the geometry of the cell can determine the direction and structure of these flows.

This paper is organized as follows. First, we perform large scale hydrodynamic simulations in ellipsoidal geometries of a fluid structure model based on recent theoretical work on cytoplasmic streaming \cite{Stein2019}. We find relaxation to vortical steady states, consistent with earlier theoretical work that found vortical steady states in spherical geometries \cite{Dutta2024}. Furthermore, we find a secondary relaxation from the initial vortical state aligned with any random direction, to a stable state in which the vortex rotates around the long axis of the ellipsoid. We call such flows, ``axisymmetric." Second, we test the prediction of axisymmetrization experimentally in stage 12 \textit{Drosophila} oocytes, which are approximately ellipsoidal. This allows us to conjecture that the flow patterns are robust and that they tend to be aligned with the long axis of the cell. Additionally, we find that the handedness of the flow is unbiased as would be expected from a symmetry breaking system. Finally, we probe the robustness of such vortical flows with a coarse grained continuum model that allows us to easily explore steady state solutions to other geometric perturbations, such as intracellular obstacles in the form of a cell nucleus.

The simplest setting for our problem is a closed domain containing a Stokes fluid and uniformly covered with motor-loaded slender filaments attached to the surface \cite{Stein2019,Stein2021} (Figure 1A). This model is based on the known biology of cytoplasmic streaming in the \textit{Drosophila} oocyte. In the late-stage oocyte, stable microtubules (MTs) are cortically anchored to the cell cortex at their minus ends \cite{Parton2011, Monteith2016}. These MTs serve as railroads for the plus-end directed nanoscale molecular motors, kinesins, that carry cellular cargos \cite{Lu2016, Lu2018}. By walking towards the free end of cortically anchored MTs, kinesin-cargo complexes exert a force on the cytoplasm, entraining the nearby fluid, and a compressive force on the inextensible MTs. For large compressive loads, the MTs can bend or buckle, leading to synergistic interactions between the deformations of the MTs and the flows they generate \cite{Civalek2011}. 

A previous analysis of this model established that when the areal density of the MTs and the force from the motors was sufficiently high, the collective interactions resulted in the emergence of a self-sustained intracellular cell-scale vortical flow in a spherical geometry \cite{Dutta2024}. The vortical flow takes the form of rigid-body rotation plus a weak bi-toroidal flow component \cite{Dutta2024}. However, oocytes are not spheres (the late-stage oocytes are well approximated by prolate ellipsoids) and cell geometry has non-trivial consequences for the emergent flows because it is non-linearly coupled to the the cytoplasmic flows \cite{Orlandini2013, Okabe2008, Pintado2017}. Geometry breaks the rotational symmetry in the problem and poses the question of how the interior flow is topologically organized. What sets the axis and chirality of the emergent flows? The present study examines these questions, which are central to understanding the potential functions of streaming flows.
\begin{figure}[ht!]
    \centering
    \includegraphics[width=\linewidth]{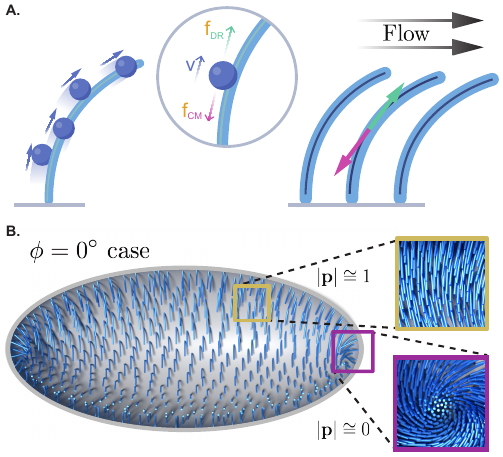}
    \caption{\textbf{Biophysical model for motor driven fluid flows:} \textbf{(A)}. Kinesin-cargo complexes (blue spheres) walk towards the free plus ends of anchored cortical MTs exerting both a drag force, $f_{DR}$, on the surrounding cytoplasm and a downwards compressive force, $f_{CM}$, on the inextensible MTs, causing them to bend. The model coarse grains the kinesin-cargo forces as a uniform line density of forces, $\sigma$. \textbf{(B)}. The orientation of microtubules in a ellipsoid during simulated flows. The flow rotates in line with the aligned MTs (yellow boxed detail) around the line connecting the defects (purple boxed detail). In ellipsoidal geometries the defects go to the poles of the cell, which corresponds to flow around the long axis of the ellipsoid. The flow is characterized by $\phi$, which measures the angle between the long axis of the ellipsoid and the axis of rotation of the flow.}
    \label{fig:my_label}
\end{figure}
\section{\label{sec:level1} Results}
\subsection{\label{sec:level2}Large scale hydrodynamic simulations in ellipsoidal geometries}
Following previous work, the joint dynamics of microtubules (MTs) coupled through a viscous fluid are described \cite{Stein2021}. The evolution of each microtubule $\mathbf{X}^{i}(s,t)$, parameterized by arc length $s$, and time $t$, evolves as:
\begin{align}
\eta (\mathbf{X}^i_{t}-\mathbf{u}) &= (\mathbf{I}+ \mathbf{X}_{s}^{i} \mathbf{X}_{s}^{i}) \cdot (\mathbf{f}^i - \sigma \mathbf{X}_{s}^{i}), 
\\
\mathbf{f}^i &= -E \mathbf{X}_{ssss}^{i} + (T^{i} \mathbf{X}_{s}^{i})_{s}.
\end{align}
The forces governing the dynamics of the fiber are $\mathbf{f}^i$, the elastic force per unit length from microtubule bending and $\sigma$, the coarse grained compressive force due to the movement of motor-cargo complexes. $E$ is the bending rigidity and $T^{i}$ is the MT tension, which acts as a Lagrange multiplier to enforce the incompressibility of each fiber, and $\eta$ is the drag coefficient. The drag coefficient on the slender fiber is given by $\eta=\frac{8 \pi \mu}{|c|}$, where $\mu$ is the viscosity and $c=\log e \epsilon^2$ is a coefficient characterizing the slenderness of the MT ($\epsilon$ is the ratio of MT length to width). Each MT fiber is clamped normally to the cortex and the free end is both force and torque-free. 

The left half of Figure 1A shows the forces on a single MT from drag and compression, and the resultant bending of the MT. The right half of Figure 1A illustrates the coarse graining of force from the motor-cargo complexes to a uniform line density of forces as well as the resultant flow field generated in the surrounding cytoplasm. The flow of the cytoplasm is governed by the forced incompressible Stokes equation and subject to a no-slip boundary condition at the cortex: 
\begin{align}
\nabla q - \mu \Delta \mathbf{u} &=  \sum\limits_{i=1}^N \int\limits_{0}^{L} \mathrm{d}s \  \mathbf{f}^{i}(\textit{s})\delta(\mathbf{x}-\mathbf{X}^{i}), \\
\nabla \cdot \bu &= 0,
\end{align}
where $q$ is pressure and $\mathbf{u}$ is the velocity field.
\begin{figure}
    \centering
    \includegraphics[width=\linewidth]{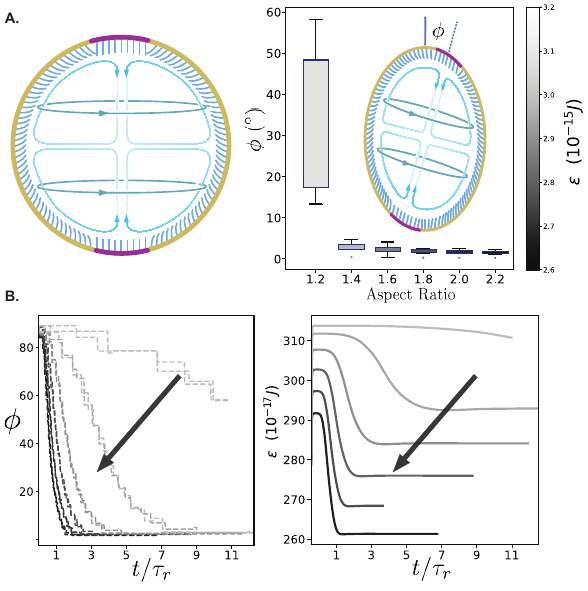} 
    \caption{\textbf{The effect of cell geometry on final flow orientation from numerical experiments:} \textbf{(A)}. The structure of flow in a spherical geometry flow is largely solid body rotation with a small bitoirodal component due to the closed geometry. Defects appear at opposite ends of the sphere (highlighted in purple). Simulations in ellipsoids of aspect ratios ($AR$) ranging from 1.2 to 2.2 robustly show that the steady state location of the defects/axis of rotation (defined by $\phi$) is consistently nearly or exactly at the poles of the ellipsoid. The elastic energy of the final state is minimized as the aspect ratio of the ellipsoid is increased.\textbf{(B)}. The location of the defect as a function of simulation time is shown for ellipsoids initialized with a steady state flow field where $\phi=90^\circ$ (left panel). $\phi$ converges to the axisymmetric flow case ($\phi=0^\circ$) for ellipsoids with aspect ratios $>1$. The convergence to the axisymmetric flow state is associated with a minimization of elastic energy (right panel). The arrow represents increasing aspect ratios ranging from 1.2 (lightest grey) - 2.2 (black).}
    \label{fig:my_label}
\end{figure}

We used large scale hydrodynamic simulations (SkellySim) to evolve fibers according to Eq. (1) and Eq. (2) in a background fluid subject to Eq. (3) and Eq. (4) in ellipsoidal geometries. The result of a representative simulation is visualized in Figure 1B. As in the case of the sphere, vortical flow emerged as the long term state of the problem. However, in the case of the ellipsoid, the axis of rigid-body rotation was independent of initial conditions and consistently close to the long axis of the ellipsoid. We characterized the flow rotation by $\phi$, the angle of inclination made by the vortical axis to the long axis of the cell. Completely aligned axisymmetric flow is characterized by $\phi=0^\circ$. 

To determine the axis of flow rotation we made use of the fact that the flow field rotates around the line connecting ``defects" in the MT fiber bed. We define ``defects" as locations on the surface where fibers are locally unaligned. Vortical flow is characterized by a largely bent or ``combed over" bed of fibers with two defects at opposite ends of the cell. Just as combing a hairy ball results in at least two topological defects, the fibers are not able to all bend over in such a way that their directions are always locally aligned, and instead two ``defects" form. An example of a defect is shown by the rightmost purple boxed detail in Figure 1B. 

To calculate the defect location we use the surface polarity field as previously done by Ref. \cite{Dutta2024}. The vector $\mathbf{p}^i(t)$ is defined as the projection of fiber $i$ onto the local tangent plane:
\begin{equation}
\mathbf{p}^i (t) = (\mathbf{I} - \mathbf{nn}) \cdot \frac{\mathbf{X}^i (L,t) - \mathbf{X}^i (0,t)}{L},
\end{equation}
where $L$ is the length of MT and $\mathbf{n}$ is the inward surface normal at $\bX^{i}(0,t)$. The surface polarity field, $\mathbf{p}(t)$ is a local average of $\mathbf{p}^{i}(t)$ vectors. When the fibers are highly bent and aligned in the same direction, $|\mathbf{p}(t)|$ is close to and bounded by 1. A representative example of locally aligned fibers is shown by the yellow boxed detail in Figure 1B. When the fibers are unaligned, $|\mathbf{p}(t)|$ is close to 0. 

Simulations from an initially straight bed of fibers relaxed to a vortical flow with two $|\mathbf{p}| \cong 0$ defects in MT order. In ellipsoidal geometries, the initial locations of the two defects were strongly dependent upon initial conditions and possibly far from alignment with the long axis. The two defects then slowly moved from their initial position of formation towards the poles of the ellipsoid, resulting in a nearly axisymmetric flow state at long times. This movement of defects from their initial position to their resting place at the poles was associated with a decrease in total elastic energy, the governing energy in this low Reynolds number, bending driven flow. The elastic energy $\mathcal{E}$, is directly proportional to the squared curvature of the fibers and is defined as:
\begin{equation}
\mathcal{E}(t) = \frac{E}{2} \sum_{i=1}^{N} \int_{0}^{L^{i}} ds [\kappa^i(s,t)]^2,
\end{equation}
where $N$ is the number of fibers and $\kappa$ is the curvature along the fiber. The elastic energy comes from the movement of molecular motors on the fibers, causing them to bend. The energy is dissipated in the form of viscous drag in the fluid flows.
\subsection{\label{sec:level2}Analysis of velocity fields from simulations in ellipsoidal geometries}

To investigate the stability of the ``axisymmetric" flow state in an ellipsoid, we performed numerical ``perturbation" experiments in which we evolved flow in a sphere until steady state. We then stretched the sphere perpendicular to the axis of rotation of flow while maintaining the same surface area and MT placement. We ensured that the boundary condition of normally clamped fibers was met. We allowed the system to evolve from this vortical state where the initial axis of rotation was $\phi=90^\circ$. The final $\phi$ was calculated and converged to $0^\circ$ as the ellipsoid became increasingly elongated (Figure 2A). This was true for ellipsoids with aspect ratios ranging from 1.2-2.2 (the aspect ratios we found in late stage \textit{Drosophila} oocytes). The elastic energy of the steady state was lowest in ellipsoids with the highest aspect ratios. 

The effective dynamics can be rationalized as follows: because the surface area remains the same, the radius of the body of the ellipsoid decreases and the MTs become closer to each other. MTs across from each other are pushing and being pushed by the flow in opposite directions; they inhibit each other more and more as the long axis of the ellipsoid narrows. This leads to a relaxation of the MTs and a consequent decrease of elastic energy. 
\begin{figure}[ht!]
\includegraphics[width=\linewidth]{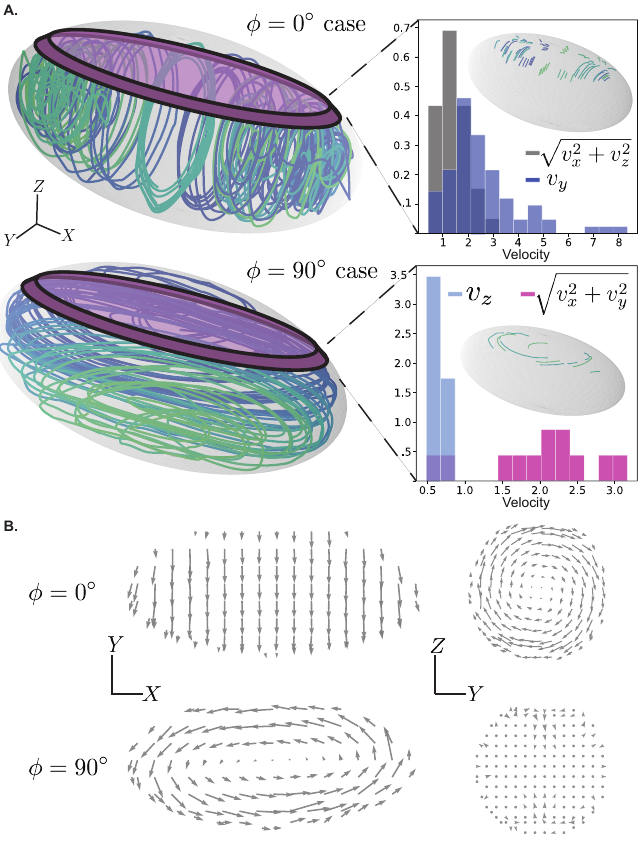}
\caption{\textbf{The effect of flow orientation on cortical velocity fields from simulation:} \textbf{(A)}. Simulated streamlines from ``axisymmetric flow" ($\phi=0^\circ$), with flow around the long axis of the ellipsoid and ``vortex flow" ($\phi=90^\circ$), with flow around the long axis of the ellipsoid. The relative magnitudes of $v_x$,$v_y$, and $v_z$ in a thin cortical section as visualized by the purple volume discriminate between the ``axisymmetric" and ``vortex" flow cases. The ability to discriminate between flow orientations has implications for live-imaging, which can only access streaklines near the cortex of the ooycte. The $X$-axis is defined by the long axis of the ellipsoid. \textbf{(B)}. The velocity field of the thin cortical slice for both $\phi=0^\circ$ and $\phi=90^\circ$. In the X-Y plane $v_y$ is dominant for axisymmetric flow, while $v_x$ is dominant in fields where $\phi=90^\circ$.}
\end{figure}

The timescale of evolution to the ``axisymmetric" state occurs on the order or $\tau_r = \frac{\eta L^4}{E}$, the single MT relaxation timescale, which is the longest timescale of the system. The two other relevant timescales in this system are $\tau_c$, the collective MT relation timescale, and $\tau_m$, the timescale associated with a motor moving a MT its own length \cite{Dutta2024}. Additionally, we found that this relaxation, which can be well fit by an exponential decay, was proportional to the aspect ratio. The higher the aspect ratio, the shorter the lifetime of the decay, $\lambda$. In other words, elongated ellipsoids were able to more quickly reach steady state (Figure 2B). 

Finally, in order to interpret experimental data, which can only be obtained from thin cortical slices of the cell, we characterized velocity streamlines on cortical slices from simulation data to predict what would be seen by a microscope for different $\phi$ values. Defining the long axis of the cell as $X$ and height as $Z$, we found that the relative magnitudes of the 3-component velocity field differed between $\phi=0^\circ$ and $\phi=90^\circ$ flow (Figure 3). When the velocity field is axisymmetric ($\phi=0^\circ$), $v_y$ always dominates. On the other hand, when the axis of rotation is perpendicular to the long axis of the cell ($\phi=90^\circ$), $v_x$ dominates.

\begin{figure*}
\includegraphics{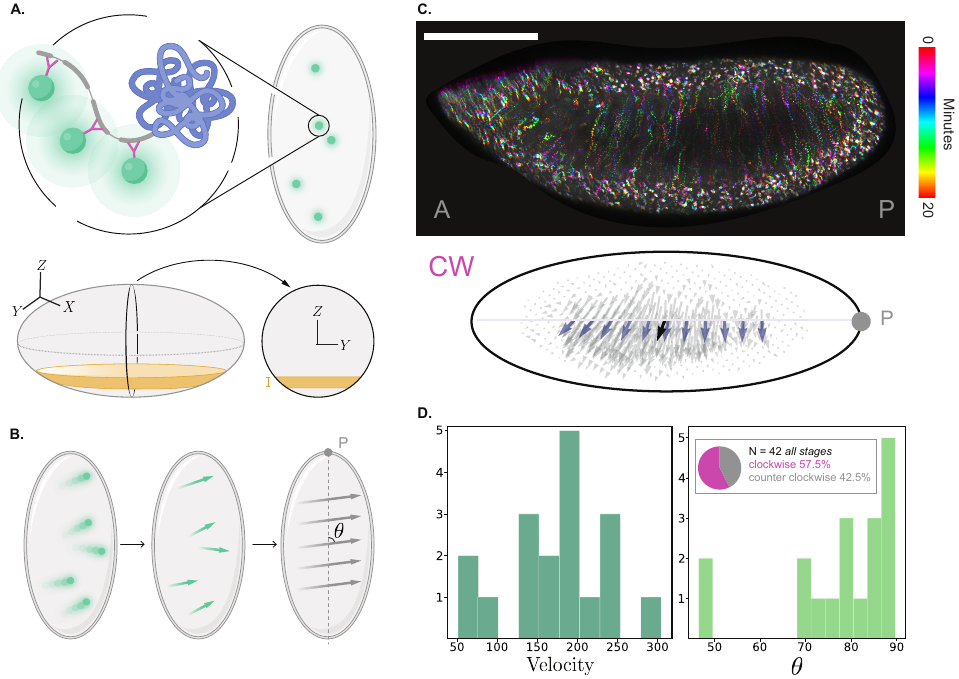}% Here is how to import EPS art
\caption{\label{fig:wide}\textbf{Live imaging of inert particles and velocity field extraction:} \textbf{(A)} Staufen proteins bound by multiple fluorescently protein tags aggregate and form inert tracer particles in stage 12 oocytes. These particles range from 0.5-2 $\mu m$ in width. Using confocal imaging, we obtained 3D cortical slices as illustrated by the region highlighted in yellow. \textbf{(B)}. The mean direction of flow, $\theta$, was extracted from a maximum projection of the 3D cortical sections with a depth of 10-15 $\mu m$. The observed flow did not change direction in an experiment or simulation over 10-15 µm, which justified using the maximum projection. We tracked particles in each frame and linked the time points to extract trajectories. The trajectories were interpolated onto a structured grid to obtain a velocity field from which the overall mean angle of flow with respect to the long axis of cell was computed. The handedness of the flow was determined with respect to the posterior end of the cell (grey dot) \textbf{(C)}. An example of the experiment/analysis pipeline used to extract the velocity flow field with a mean direction, represented by the black arrow. The posterior of the cell is highlighted by the grey dot and ``P”; scale bar = 100 $\mu m$. \textbf{(D)}. In stage 12 oocytes, the distribution of maximum measured velocity is higher than that of stage 10 oocytes \cite{Dutta2024}. Across all stages of streaming (10-12), 57.5 percent rotated CW and 42.5 percent rotated CCW. The distribution of mean flow angles peaked around $\theta$ = 90. This distribution is consistent with 3D flow fields where the defects are close to or at the poles ($\phi = 0^\circ$).}
\end{figure*}
\subsection{\label{sec:level2} Live imaging and particle tracking}

Next, we turned to experiments to test the predictions of the numerical analysis: (1) that cell geometry plays a role in setting the direction of the flow field in \textit{Drosophila} oocytes, and (2) the sense of rotation of the flow should be unbiased.While experiments observing flows in \textit{Drosophila} oocytes date back to the 1980's, there has been no characterization of their handedness or axis of rotation \cite{Gutzeit1982,Quinlan2016}. 

Our experiments differ from previous experiments characterizing cytoplasmic streaming in \textit{Drosophila} oocytes in three ways. Firstly, we focused on late-stage oocytes where the cell is approximately ellipsoidal. In late-stage oocytes, the ”dumping” process that delivers cytoplasm to the oocyte from the attached sister nurse cells is complete \cite{Quinlan2016}. Flows in \textit{Drosophila} oocytes can be broadly characterized as occurring before dumping, during dumping, and after dumping. Before dumping (stage 10), the oocyte geometry is complex and ill-approximated by any simple geometry such as a ellipsoid. During dumping (stage 11), the oocyte changes volume at a rate comparable to the velocity of the flow field and there is an additional and significant input of volume from the nurse cells. This contribution of nurse cell flows renders the dynamics more complex than we currently model \cite{Alsous2021}. However, after dumping (stage 12), the flow from the nurse cells stops and the volume of the oocyte ceases to change rapidly with respect to the velocity of the flows. At stage 12, the oocyte geometry is well approximated by a ellipsoid. By focusing on flow during this stage, we therefore avoided the confounding effects from dumping flows, complicated geometries, and any early flows which may not have reached steady state. Instead, stage 12 flow provides an excellent approximation to test our numerical predictions.

Secondly, we performed the first experiments in which single particle trajectories are visualized in the flow.  To do so, we took advantage of a \textit{Drosophila} strain expressing Staufen-SunTag \cite{Lu2016} (Figure 4.A). Staufen, an RNA-binding protein, is bound by multiple GFP-tagged nanobodies (SunTag), which results in a tendency to aggregate and create colloidal clumps of Staufen-SunTag molecules in the oocyte, with radii ranging from 500 $nm$ - 2 $\mu m$. This generates fluorescently labeled particles throughout the oocyte cytoplasm that are advected by the flow and can be used as molecular tracers.

And lastly, we developed analytical methods to extract both individual particle tracks and velocity profiles, as well as characterize the flow orientation and handedness. We performed live confocal imaging of stage 12 oocytes expressing Staufen-SunTag \cite{Lu2018}. Because the flow fields do not change dramatically slice to slice, the 10-15 $\mu m$ stack of images was projected to maximize the available data. The particle tracking produced individual trajectories for each tracked particle (Figure 4B). 

Based on the observation that the trajectories tend to be aligned and perpendicular to the long axis, we used the individual tracks to extract a velocity field from which we could get an average direction of flow. To do so we made use of our knowledge about the flow field to interpolate the trajectories onto a structured grid. Because we observed that the fluid did not change directions locally and because the fluid is incompressible, we assumed that the flow of a particle between two parallel trajectories is parallel and in the same direction. We thereby averaged information from the nearest neighbors of the known particle trajectories to points on a structured grid to generate an interpolated velocity field (Figure 4B and 4C). We observed that the velocity fields were largely vortex free and that the average angle of flow was perpendicular to the long axis of the cell, which we defined by fitting a ellipsoid to the cell boundary.

To analyze the flow field, we characterized the velocity field by a single parameter: average angle of flow with respect to the long axis of the cell, $\theta$ (Figure 4B). We chose this method because the particle tracks were largely perpendicular to the long axis of the cell for nearly every oocyte imaged. We excluded examples with an obvious vortex by calculating the maximum curl of each velocity field and excluding examples that have points with a curl value above a threshold (there was only one such case). 

Our simulations predicted that the axis of flow rotation should be aligned with the long axis of the oocyte. Our experimental results showed that the average angle of flow was peaked at $90^\circ$, which is consistent with flow rotating about the long axis of the cell. With the exception of one stage 12 oocyte (of 18 in total) in which flow appeared as a vortex, the measured flow fields were consistent with axisymmetric flow, as seen by the histogram in Figure 4D. Additionally, we measured the maximum velocity along the axisymmetric direction and found that on average, the maximum velocity of solid body rotation was 180 $nm/s$. This is substantially higher than the velocities measured by in stage 10B oocytes, which had a maximum speed of ~100 $nm/s$ \cite{Dutta2024} . 

Likewise, our symmetry breaking model and simulations predicted no bias in the handedness of the flows. From experiments, it was equally likely for a flow to be clockwise as it was counterclockwise (Figure 3D).
\begin{figure}[!ht]
    \centering
    \includegraphics[width=\linewidth]{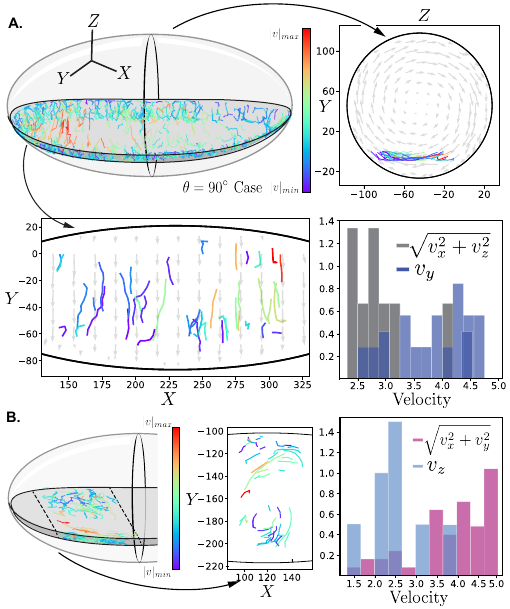}
    \caption{\label{fig:wide}\textbf{Differentiating flow orientation from 3D single particle tracks:} \textbf{(A)}. 3D particle tracks from a live imaging experiment with measured $\theta \cong 90^\circ$. The streamlines from both the short axis and long axis perspective are overlaid on an axisymmetric velocity field as computed by simulation ($\phi=0^\circ$). The $x$-axis is set by the long axis of the oocyte. The relative magnitudes of the streamlines from a cortical 3D volume with average angle of flow $\theta \cong 90^\circ$, are consistent with those from simulated axisymmetric flow ($\phi=0^\circ$) as seen in Figure 1. \textbf{(B)}. 3D particle tracks from the single experiment with a visible vortex. The distribution of velocity components is markedly different from the axisymmetric case and again matches the prediction from a simulation where $\phi=90^\circ$.}

\end{figure}

We used the full 3D particle tracks to compare the imaged flow field with the simulated ones. Analysis of the streamlines of simulated flow fields revealed a pronounced difference in the relative magnitudes of the velocity components, $v_x$,$v_y$,and $v_z$, where the $X$-axis was defined by the long axis of the ellipsoid. If the flow field was close to axisymmetric, streamlines on cortical slices had $v_y>v_z,v_x$. If the flow field rotated perpendicular to the long axis of the cell, then the streamlines had $v_x,v_y>v_z$ (Figure 3). Figure 5 shows that the analysis of experimental streamline velocity in two cases, axisymmetric and perpendicular to axisymmetric, is consistent with the predictions of simulated flow fields. In other words, measured flows with an average $\theta \cong 90^\circ$ are consistent with simulations of ``axisymmetric" flows where $\phi=0^\circ$. 

\subsection{\label{sec:level2} Robustness to perturbation using coarse grained continuum model}
To summarize our results so far, we found that flows can withstand geometric perturbations in the shape of the domain and that the axis of rotation is biased to be around the long axis. Another type of perturbation arises from internal structures, such as organelles. Though the geometry of the oocyte can be well approximated by an ellipsoid, the interior of the oocyte is densely packed with yolk granules, Golgi, endoplasmic reticulum, and various other cellular components. The largest of these components is the nucleus, with a radius of approximately $10 \mu m$, located near the anterior of the oocyte (Figure 6B) \cite{Alsous2021,Quinlan2016,Lepesant2024}. It is thus natural to ask: how robust are the collective dynamics to large interior geometric perturbations that could alter the topology of the emergent flows?

This question can be most readily addressed using a recently proposed coarse-grained 
``active carpet" model where we model the nucleus as a fixed no-slip surface \cite{chakrabarti2024cytoplasmic}. In contrast to evolving the dynamics of individual MT filaments, the active carpet theory evolves a surface polarity field that quantifies the mean orientation of the MTs at a point. The model takes the form of a boundary force field coupled to an internal Stokesian flow and can be easily adapted to a range of self-organized, cortex-driven intracellular flows \cite{chakrabarti2024cytoplasmic}.
\begin{figure*}[!ht]
    \centering
    \includegraphics{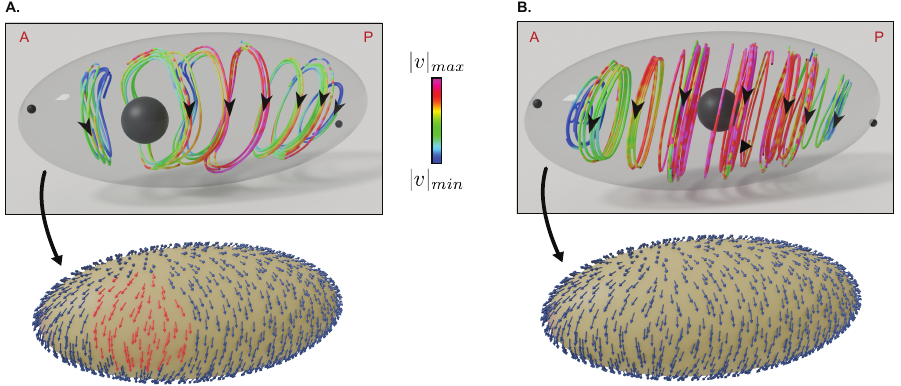}
    \caption{\label{fig:wide} \textbf{The role of geometric obstacles in emergent flow topology:} \textbf{(A)} Streamlines from the emergent steady state of the twister flow in the presence of a nucleus. The nucleus is located on the posterior side close to the MT bed. The color code of the streamlines indicates the magnitude of the fluid velocity and the arrow indicates the direction of circulation. We observe that the fluid flow slows down near the nucleus due to the no-slip boundary condition. The associated polarity field is shown below: (left) same view angle as the twister flow reveals splaying of the MT polarity near the nucleus; (right) A rotated view shows the positioning of +1 topological defect. The anterior (A) and posterior (P) poles are labeled for convenience and the defect locations are depicted by the black spheres.  \textbf{(B)} The same is shown for a similar-sized geometric obstacle placed at the center of the egg cell. Unlike \textbf{(A)}, the flow does not break AP symmetry and retains its robust almost axisymmetric structure. The streaming speed $U_s$ for these calculations range between $200-300$ nm/s.}

\end{figure*}

In this coarse-grained model, we characterize the dense MT carpet by a uniform areal density $\bar{\rho}$ and its orientation by a surface polarity field $\mathbf{P}(\by,t)$, where $\by$ is the surface coordinate on the cell boundary. The polarity
field evolves according to Jeffery’s equation as
\begin{equation}\label{eq:polarity}
    \partial_t \bP = (\bI - \bP \bP) \cdot \nabla \bu\big|_{\partial D} \cdot \bP + \bT_0 \times \bP.
\end{equation}
Here $\bu(\bx,t)$ is the internal cytoplasmic fluid velocity, $\partial D$ is the cell boundary, and $\bT_0$ is a restoring torque that mimics the bending response of clamped filament by trying to align the microtubules with the local inward normal to the surface. The cytoplasmic fluid velocity is determined by solving the incompressible Stokes equation subject to the no-slip boundary condition on the cell surface. The activity of motor proteins driving the internal flow is coarse-grained as surface stress jump $\bar{\mathbf{f}}(\by,t)$ across an interface $\partial S:= \partial D +  \bn(\by)$ formed by the tips of the MT vectors. In the dimensionless form, the momentum balance for the fluid flow reads
\begin{alignat}{2}
    -\nabla q + \Delta \bu = \mathbf{0}, &  \ \ \nabla \cdot \bu  = 0, \label{eq:homst}& \\ 
    \bu(\by,t) = \mathbf{0}, &  \ \ \text{ on } \partial D, & \\ 
    \llbracket \bu \rrbracket = \mathbf{0}, \ \ \llbracket \boldsymbol{\sigma}  \cdot \bn \rrbracket (\by,t) = -\bar{\bff}(\by,t).& \ \ \text{ on } \partial S. & \label{eq:stokeappx}
\end{alignat}
Here $\llbracket a \rrbracket = a|_{\partial S^+} - a|_{\partial S^-}$ denotes the jump of any variable across the interface $\partial S$, and $\boldsymbol{\sigma} = -q \bI + (\nabla \bu + \nabla \bu^T)$ is the Newtonian stress tensor. The surface stress jump is given as
\begin{equation}\label{eq:jump}
    \bar{\bff}(\by,t) =  \bar{\rho} \bar{\sigma} \mathbf{P} + \bar{\rho} \chi \left[ \mathbf{T}_0 \times \mathbf{P}-\frac{\mathbf{P P P}}{2}: \nabla \mathbf{u}\Big|_{\partial D} \right]. 
\end{equation}
The first term in Eq.~\eqref{eq:jump} arises from the forces exerted by the Kinesin-1 motor proteins on the fluid and is proportional to the density $\bar{\rho}$ and forcing from individual motor proteins $\bar{\sigma}$. The second term represents forces stemming both from the torsional spring and the rod's response in the mean-field fluid flow \cite{chakrabarti2024cytoplasmic}; $\chi$ is a geometric parameter that depends on the aspect ratio of the MTs. Taken together, this model integrates a boundary force field as defined by Eq. ~\eqref{eq:jump} to an internal homogeneous Stokes flow delineated in Eqs. \eqref{eq:homst}-\eqref{eq:stokeappx}. The forcing and the emergent flow evolve self-consistently with a PDE that details the polarity of the MT bed as laid out in Eq.~\eqref{eq:polarity}.

Figure 6 illustrates the steady-state flows computed using this coarse-grained model. While the global structure of the flow remained fairly robust, it now exhibits key differences from the cases without a nucleus. First, the flows are no longer axisymmetric due to the asymmetric positioning of the nucleus and the no-slip boundary condition on its surface. Secondly, we observed that the slowing down of the flows around the nucleus was accompanied by splaying of the surface polarity field in the MT orientation, as depicted in Figure 6B. Finally, the anterior-posterior symmetry breaking of the oocyte had interesting consequences for the weaker bitoroidal flow. In simulations with unperturbed prolate ellipsoids, the mid-plane of the oocyte served as a plane of symmetry that separated the recirculating regions arising from the bitoroidal flow. In contrast, here the nucleus plays the role of a geometric separatrix of the toroidal flows on two asymmetric halves of the oocyte. These flows are topologically distinct and do no mix with one another. We believe this could have significant implications in the context of transport and mixing of cytoplasmic contents. In conclusion, the results of the active carpet model provide further evidence that cytoplasmic streaming is robust to certain geometric perturbations. Though the local structure of the flow might be affected, the overall flow retains its axisymmetric and vortical structure.

\section{\label{sec:level1} Discussion}
There is relatively little understood about how the geometry of the cell affects the processes within it. We study this question in the important context of self organized cytoplasmic streaming in the \textit{Drosophila} oocyte. In this work, we have used a combination of large scale simulations, live imaging experiments, and reduced models of cytoplasmic streaming. 

Our hydrodynamics simulations make a prediction about the role of geometry in the real system, which we then confirmed experimentally. While observations of cytoplasmic streaming have been around since the 1980s, there has been no quantitative characterization of flows in late stage oocytes. In earlier stage oocytes, previous studies relied on particle image velocimetry (PIV). We developed a single particle tracking pipeline for tracking and visualizing inert particles in the flow. We extracted information about the average angle of the flow and handedness from max projections of the 3D time series. Finally, we were able to use the full 3D particle tracks to compare predictions from simulations and distinguish flow orientations, even from thin cortical slices. Our experimental findings matched our computational predictions: self organized flow emerges and rotates consistently around the long axis of the cell with no bias in handedness. 

How a system evolves from an initial configuration with many possible solutions to selecting a single solution is a challenging problem. However, prior to this work, it was unclear how topology could contribute to the loss of solutions in this specific system. Numerically and experimentally, we argue that the domain of the container aids in reducing the number of possible flow orientations to just one: flow around the long axis. Results from the model are realized in the actual cell. Just as the simulations provide insight about the nature of the flow, the experimental results, too, provide insight about the accuracy of the simulations and underlying model. By showing that, just as simulations predict, there is a preferred direction of flow in real oocytes, we have strengthened both the evidence for the current biophysical model and our understanding of boundary driven flows in oocytes. Our updated picture of cytoplasmic streaming paves the way for future work to characterize the functional properties of the flow, namely how it transports molecules and mixes yolk granules.

\section{\label{sec:level1} Materials and Methods}
\subsection{\label{sec:level2} Fly stocks and genetics}

Flies were raised on standard cornmeal fly food prepared by Princeton University, which was  supplemented with active yeast 24-48 hours before dissection. The stocks were kept at $25^\circ$ C. The standard UAS system for targeted gene expression in \textit{Drosophila} was used to visualize Staufen protein aggregates in the cytoplasmic flows. \textit{UASp-scFv; UASp-Staufen-SunTag} flies \cite{Lu2018} co-expressing scFv tagged GFP and SunTag tagged Staufen proteins were crossed to nos driver flies, \textit{GAL4-nos.NGT} \cite{tracey2000quantitative}. Staufen is recruited to the posterior in wildtype oocytes. However, in stage 12 oocytes, Staufen did not visibly bind to the posterior end of the imaging plane; thus fluorescent Staufen aggregates were used as an inert tracer particles. 

\subsection{\label{sec:level2} Live imaging}
Female flies of the correct genotype were anesthetized on CO$_2$ pads and dissected in Schneider's Mix media supplemented with Insulin, FBS, and streptomycin/penicilin, and pH adjusted as described by \cite{prasad2007protocol}. The egg chambers were imaged in MatTek 35mm glass bottom culture dishes 
using a Nikon AX laser scanning confocal microscope and the NIS-Elements software at the CCB Scope Observatory at the Flatiron Institute. Imaging was performed using a 40x/1.2 silicon oil objective lens. Pinhole settings ranged from 1.0 to 1.2 Airy units. Excitation of the fluorophore, GFP, was performed at 489 nm. 3D volumes 15-20 $\mu$m deep were acquired every 10-20 seconds at 1 $\mu$m/step. 
 
\subsection{\label{sec:level2} Image analysis and velocity field reconstruction}

Fiji's TrackMate plugin was used to perform single particle tracking on 2D max projection slices from 10-15 $\mu m$ depth 3D volumes. The max projection was used to increase the resolution and was justified by the fact that the velocity field did not vary substantially in thin 10-15 $\mu m$ cortical sections. The degree of gaussian (DoG) filter was used to detect particles of diameter ~1.5-2.5 $\mu m$, which corresponded to the average size of the imaged GFP-Staufen tracer particles. A Linear Assignment Problem (LAP) tracker was used to connect particle trajectories and then filtered for trajectories of a minimum length. The average velocity fields were acquired by integrating information across the time lapse images. Because the flow fields were relatively stable in Stage 12 oocytes, a steady state flow was assumed. Each time point of a particle trajectory resulted in a local velocity vector at the instantaneous position of the particle. Velocity vectors from each time point in the movie were used to generate a velocity field. To interpolate the velocity field onto a structured grid, each vector within a 5 $\mu m$ radius of a point on the structured grid was averaged. If there were fewer than 30 vectors within 5 $\mu m$, the radius of search was increased until there were 30 velocity vectors within the circle.

A least squares method was used to fit an ellipse with semi-major axis, $A$, and semi-minor axis, $B$, to the boundary of the oocyte. The ellipse long axis of the ellipse defined the anterior-posterior axis. To account for the fact that the edge of the oocyte is coated by a microtubule bed, only data within 3/4th of the semi-major and semi-minor axis were considered. The interpolated velocity field was binned along the length of the long axis and the mean vector was calculated. The angle, $\theta$, was reported with respect to the long axis of the cell. The handedness of the flow was counted with respect to the posterior of the cell.

We were careful in correlating velocity profiles from image slices near the cortex to entire flow fields in the oocyte because we were dealing with an ill-posed inverse problem. Nearly identical slices near the cortex with similar $\theta$'s could correspond to different flow fields in the body. Our experimental result is consistent with sampling from axisymmetric or close to axisymmetric flow orientations. Therefore the distribution of $\theta$'s from Figure 3 in which $\theta$ is peaked around $90^\circ$ indicates a predominance of near ``axisymmetric" flows, just as predicted by simulation (Figure 2).

3D particle tracks were extracted using the DoG filter and nearest neighbor algorithm in Fiji's TrackMate plugin. Ellipses were fit using least squares to every slice in the z-stack and the resulting stack of ellipses and their known z-spacing was used to fit the entire oocyte to an ellipsoid. The long axis of the ellipsoid was set as the X-axis and the direction of the z-stack was set as the Z-axis. Within this coordinate system, the 3D streaklines were used to obtain information about the relative magnitudes particle velocity components,  $v_x, \ v_y$, and $v_z$.

\begin{acknowledgments}
We wish to acknowledge David Stein, Wen Lu, Vladimir Gelfand, Manas Rachh, Jasmin Imran Alsous, and Howard Stone for valuable discussions. We wish to thank Wen Lu and Vladimir Gelfand for their generous contribution of the fly lines used in all experiments.
\end{acknowledgments}

\bibliography{maintext}% Produces the bibliography via BibTeX.

\begin{thebibliography}{20}%
\makeatletter
\providecommand \@ifxundefined [1]{%
 \@ifx{#1\undefined}
}%
\providecommand \@ifnum [1]{%
 \ifnum #1\expandafter \@firstoftwo
 \else \expandafter \@secondoftwo
 \fi
}%
\providecommand \@ifx [1]{%
 \ifx #1\expandafter \@firstoftwo
 \else \expandafter \@secondoftwo
 \fi
}%
\providecommand \natexlab [1]{#1}%
\providecommand \enquote  [1]{``#1''}%
\providecommand \bibnamefont  [1]{#1}%
\providecommand \bibfnamefont [1]{#1}%
\providecommand \citenamefont [1]{#1}%
\providecommand \href@noop [0]{\@secondoftwo}%
\providecommand \href [0]{\begingroup \@sanitize@url \@href}%
\providecommand \@href[1]{\@@startlink{#1}\@@href}%
\providecommand \@@href[1]{\endgroup#1\@@endlink}%
\providecommand \@sanitize@url [0]{\catcode `\\12\catcode `\$12\catcode
  `\&12\catcode `\#12\catcode `\^12\catcode `\_12\catcode `\%12\relax}%
\providecommand \@@startlink[1]{}%
\providecommand \@@endlink[0]{}%
\providecommand \url  [0]{\begingroup\@sanitize@url \@url }%
\providecommand \@url [1]{\endgroup\@href {#1}{\urlprefix }}%
\providecommand \urlprefix  [0]{URL }%
\providecommand \Eprint [0]{\href }%
\providecommand \doibase [0]{https://doi.org/}%
\providecommand \selectlanguage [0]{\@gobble}%
\providecommand \bibinfo  [0]{\@secondoftwo}%
\providecommand \bibfield  [0]{\@secondoftwo}%
\providecommand \translation [1]{[#1]}%
\providecommand \BibitemOpen [0]{}%
\providecommand \bibitemStop [0]{}%
\providecommand \bibitemNoStop [0]{.\EOS\space}%
\providecommand \EOS [0]{\spacefactor3000\relax}%
\providecommand \BibitemShut  [1]{\csname bibitem#1\endcsname}%
\let\auto@bib@innerbib\@empty
%</preamble>
\bibitem [{\citenamefont {Bialecki}\ \emph {et~al.}(2017)\citenamefont
  {Bialecki}, \citenamefont {Kazmierczak},\ and\ \citenamefont
  {Lipniacki}}]{Bialecki2017}%
  \BibitemOpen
  \bibfield  {author} {\bibinfo {author} {\bibfnamefont {S.}~\bibnamefont
  {Bialecki}}, \bibinfo {author} {\bibfnamefont {B.}~\bibnamefont
  {Kazmierczak}},\ and\ \bibinfo {author} {\bibfnamefont {T.}~\bibnamefont
  {Lipniacki}},\ }\bibfield  {title} {\bibinfo {title} {Polarization of concave
  domains by traveling wave pinning},\ }\bibfield  {journal} {\bibinfo
  {journal} {PLoS ONE}\ }\textbf {\bibinfo {volume} {12}},\ \href
  {https://doi.org/10.1371/journal.pone.0190372} {10.1371/journal.pone.0190372}
  (\bibinfo {year} {2017})\BibitemShut {NoStop}%
\bibitem [{\citenamefont {Pintado}\ \emph {et~al.}(2017)\citenamefont
  {Pintado}, \citenamefont {Sampaio}, \citenamefont {Tavares}, \citenamefont
  {Montenegro-Johnson}, \citenamefont {Smith},\ and\ \citenamefont
  {Lopes}}]{Pintado2017}%
  \BibitemOpen
  \bibfield  {author} {\bibinfo {author} {\bibfnamefont {P.}~\bibnamefont
  {Pintado}}, \bibinfo {author} {\bibfnamefont {P.}~\bibnamefont {Sampaio}},
  \bibinfo {author} {\bibfnamefont {B.}~\bibnamefont {Tavares}}, \bibinfo
  {author} {\bibfnamefont {T.~D.}\ \bibnamefont {Montenegro-Johnson}}, \bibinfo
  {author} {\bibfnamefont {D.~J.}\ \bibnamefont {Smith}},\ and\ \bibinfo
  {author} {\bibfnamefont {S.~S.}\ \bibnamefont {Lopes}},\ }\bibfield  {title}
  {\bibinfo {title} {Dynamics of cilia length in left–right development},\
  }\bibfield  {journal} {\bibinfo  {journal} {Royal Society Open Science}\
  }\textbf {\bibinfo {volume} {4}},\ \href
  {https://doi.org/10.1098/rsos.161102} {10.1098/rsos.161102} (\bibinfo {year}
  {2017})\BibitemShut {NoStop}%
\bibitem [{\citenamefont {Quinlan}(2016)}]{Quinlan2016}%
  \BibitemOpen
  \bibfield  {author} {\bibinfo {author} {\bibfnamefont {M.~E.}\ \bibnamefont
  {Quinlan}},\ }\href {https://doi.org/10.1146/annurev-cellbio-111315-125416}
  {\bibinfo {title} {Cytoplasmic streaming in the drosophila oocyte}} (\bibinfo
  {year} {2016})\BibitemShut {NoStop}%
\bibitem [{\citenamefont {Dutta}\ \emph {et~al.}(2024)\citenamefont {Dutta},
  \citenamefont {Farhadifar}, \citenamefont {Lu} \emph {et~al.}}]{Dutta2024}%
  \BibitemOpen
  \bibfield  {author} {\bibinfo {author} {\bibfnamefont {S.}~\bibnamefont
  {Dutta}}, \bibinfo {author} {\bibfnamefont {R.}~\bibnamefont {Farhadifar}},
  \bibinfo {author} {\bibfnamefont {W.}~\bibnamefont {Lu}}, \emph {et~al.},\
  }\bibfield  {title} {\bibinfo {title} {Self-organized intracellular
  twisters},\ }\href {https://doi.org/10.1038/s41567-023-02372-1} {\bibfield
  {journal} {\bibinfo  {journal} {Nat. Phys.}\ }\textbf {\bibinfo {volume}
  {20}},\ \bibinfo {pages} {666} (\bibinfo {year} {2024})}\BibitemShut
  {NoStop}%
\bibitem [{\citenamefont {Forrest}\ and\ \citenamefont
  {Gavis}(2003)}]{Forrest2003}%
  \BibitemOpen
  \bibfield  {author} {\bibinfo {author} {\bibfnamefont {K.~M.}\ \bibnamefont
  {Forrest}}\ and\ \bibinfo {author} {\bibfnamefont {E.~R.}\ \bibnamefont
  {Gavis}},\ }\bibfield  {title} {\bibinfo {title} {Live imaging of endogenous
  rna reveals a diffusion and entrapment mechanism for nanos mrna localization
  in drosophila},\ }\bibfield  {journal} {\bibinfo  {journal} {Current
  Biology}\ }\textbf {\bibinfo {volume} {13}},\ \href
  {https://doi.org/10.1016/S0960-9822(03)00451-2}
  {10.1016/S0960-9822(03)00451-2} (\bibinfo {year} {2003})\BibitemShut
  {NoStop}%
\bibitem [{\citenamefont {Stein}\ and\ \citenamefont
  {Shelley}(2019)}]{Stein2019}%
  \BibitemOpen
  \bibfield  {author} {\bibinfo {author} {\bibfnamefont {D.~B.}\ \bibnamefont
  {Stein}}\ and\ \bibinfo {author} {\bibfnamefont {M.~J.}\ \bibnamefont
  {Shelley}},\ }\bibfield  {title} {\bibinfo {title} {Coarse graining the
  dynamics of immersed and driven fiber assemblies},\ }\bibfield  {journal}
  {\bibinfo  {journal} {Physical Review Fluids}\ }\textbf {\bibinfo {volume}
  {4}},\ \href {https://doi.org/10.1103/PhysRevFluids.4.073302}
  {10.1103/PhysRevFluids.4.073302} (\bibinfo {year} {2019})\BibitemShut
  {NoStop}%
\bibitem [{\citenamefont {Stein}\ \emph {et~al.}(2021)\citenamefont {Stein},
  \citenamefont {Canio}, \citenamefont {Lauga}, \citenamefont {Shelley},\ and\
  \citenamefont {Goldstein}}]{Stein2021}%
  \BibitemOpen
  \bibfield  {author} {\bibinfo {author} {\bibfnamefont {D.~B.}\ \bibnamefont
  {Stein}}, \bibinfo {author} {\bibfnamefont {G.~D.}\ \bibnamefont {Canio}},
  \bibinfo {author} {\bibfnamefont {E.}~\bibnamefont {Lauga}}, \bibinfo
  {author} {\bibfnamefont {M.~J.}\ \bibnamefont {Shelley}},\ and\ \bibinfo
  {author} {\bibfnamefont {R.~E.}\ \bibnamefont {Goldstein}},\ }\bibfield
  {title} {\bibinfo {title} {Swirling instability of the microtubule
  cytoskeleton},\ }\bibfield  {journal} {\bibinfo  {journal} {Physical Review
  Letters}\ }\textbf {\bibinfo {volume} {126}},\ \href
  {https://doi.org/10.1103/PhysRevLett.126.028103}
  {10.1103/PhysRevLett.126.028103} (\bibinfo {year} {2021})\BibitemShut
  {NoStop}%
\bibitem [{\citenamefont {Parton}\ \emph {et~al.}(2011)\citenamefont {Parton},
  \citenamefont {Hamilton}, \citenamefont {Ball}, \citenamefont {Yang},
  \citenamefont {Cullen}, \citenamefont {Lu}, \citenamefont {Ohkura},\ and\
  \citenamefont {Davis}}]{Parton2011}%
  \BibitemOpen
  \bibfield  {author} {\bibinfo {author} {\bibfnamefont {R.~M.}\ \bibnamefont
  {Parton}}, \bibinfo {author} {\bibfnamefont {R.~S.}\ \bibnamefont
  {Hamilton}}, \bibinfo {author} {\bibfnamefont {G.}~\bibnamefont {Ball}},
  \bibinfo {author} {\bibfnamefont {L.}~\bibnamefont {Yang}}, \bibinfo {author}
  {\bibfnamefont {C.~F.}\ \bibnamefont {Cullen}}, \bibinfo {author}
  {\bibfnamefont {W.}~\bibnamefont {Lu}}, \bibinfo {author} {\bibfnamefont
  {H.}~\bibnamefont {Ohkura}},\ and\ \bibinfo {author} {\bibfnamefont
  {I.}~\bibnamefont {Davis}},\ }\bibfield  {title} {\bibinfo {title} {A
  par-1-dependent orientation gradient of dynamic microtubules directs
  posterior cargo transport in the drosophila oocyte},\ }\bibfield  {journal}
  {\bibinfo  {journal} {Journal of Cell Biology}\ }\textbf {\bibinfo {volume}
  {194}},\ \href {https://doi.org/10.1083/jcb.201103160}
  {10.1083/jcb.201103160} (\bibinfo {year} {2011})\BibitemShut {NoStop}%
\bibitem [{\citenamefont {Monteith}\ \emph {et~al.}(2016)\citenamefont
  {Monteith}, \citenamefont {Brunner}, \citenamefont {Djagaeva}, \citenamefont
  {Bielecki}, \citenamefont {Deutsch},\ and\ \citenamefont
  {Saxton}}]{Monteith2016}%
  \BibitemOpen
  \bibfield  {author} {\bibinfo {author} {\bibfnamefont {C.~E.}\ \bibnamefont
  {Monteith}}, \bibinfo {author} {\bibfnamefont {M.~E.}\ \bibnamefont
  {Brunner}}, \bibinfo {author} {\bibfnamefont {I.}~\bibnamefont {Djagaeva}},
  \bibinfo {author} {\bibfnamefont {A.~M.}\ \bibnamefont {Bielecki}}, \bibinfo
  {author} {\bibfnamefont {J.~M.}\ \bibnamefont {Deutsch}},\ and\ \bibinfo
  {author} {\bibfnamefont {W.~M.}\ \bibnamefont {Saxton}},\ }\bibfield  {title}
  {\bibinfo {title} {A mechanism for cytoplasmic streaming: Kinesin-driven
  alignment of microtubules and fast fluid flows},\ }\bibfield  {journal}
  {\bibinfo  {journal} {Biophysical Journal}\ }\textbf {\bibinfo {volume}
  {110}},\ \href {https://doi.org/10.1016/j.bpj.2016.03.036}
  {10.1016/j.bpj.2016.03.036} (\bibinfo {year} {2016})\BibitemShut {NoStop}%
\bibitem [{\citenamefont {Lu}\ \emph {et~al.}(2016)\citenamefont {Lu},
  \citenamefont {Winding}, \citenamefont {Lakonishok}, \citenamefont
  {Wildonger},\ and\ \citenamefont {Gelfand}}]{Lu2016}%
  \BibitemOpen
  \bibfield  {author} {\bibinfo {author} {\bibfnamefont {W.}~\bibnamefont
  {Lu}}, \bibinfo {author} {\bibfnamefont {M.}~\bibnamefont {Winding}},
  \bibinfo {author} {\bibfnamefont {M.}~\bibnamefont {Lakonishok}}, \bibinfo
  {author} {\bibfnamefont {J.}~\bibnamefont {Wildonger}},\ and\ \bibinfo
  {author} {\bibfnamefont {V.~I.}\ \bibnamefont {Gelfand}},\ }\bibfield
  {title} {\bibinfo {title} {Microtubule-microtubule sliding by kinesin-1 is
  essential for normal cytoplasmic streaming in drosophila oocytes},\
  }\bibfield  {journal} {\bibinfo  {journal} {Proceedings of the National
  Academy of Sciences of the United States of America}\ }\textbf {\bibinfo
  {volume} {113}},\ \href {https://doi.org/10.1073/pnas.1522424113}
  {10.1073/pnas.1522424113} (\bibinfo {year} {2016})\BibitemShut {NoStop}%
\bibitem [{\citenamefont {Lu}\ \emph {et~al.}(2018)\citenamefont {Lu},
  \citenamefont {Lakonishok}, \citenamefont {Serpinskaya}, \citenamefont
  {Kirchenbüechler}, \citenamefont {Ling},\ and\ \citenamefont
  {Gelfand}}]{Lu2018}%
  \BibitemOpen
  \bibfield  {author} {\bibinfo {author} {\bibfnamefont {W.}~\bibnamefont
  {Lu}}, \bibinfo {author} {\bibfnamefont {M.}~\bibnamefont {Lakonishok}},
  \bibinfo {author} {\bibfnamefont {A.~S.}\ \bibnamefont {Serpinskaya}},
  \bibinfo {author} {\bibfnamefont {D.}~\bibnamefont {Kirchenbüechler}},
  \bibinfo {author} {\bibfnamefont {S.~C.}\ \bibnamefont {Ling}},\ and\
  \bibinfo {author} {\bibfnamefont {V.~I.}\ \bibnamefont {Gelfand}},\
  }\bibfield  {title} {\bibinfo {title} {Ooplasmic flow cooperates with
  transport and anchorage in drosophila oocyte posterior determination},\
  }\bibfield  {journal} {\bibinfo  {journal} {Journal of Cell Biology}\
  }\textbf {\bibinfo {volume} {217}},\ \href
  {https://doi.org/10.1083/JCB.201709174} {10.1083/JCB.201709174} (\bibinfo
  {year} {2018})\BibitemShut {NoStop}%
\bibitem [{\citenamefont {Ömer Civalek}\ and\ \citenamefont {Çiĝdem
  Demir}(2011)}]{Civalek2011}%
  \BibitemOpen
  \bibfield  {author} {\bibinfo {author} {\bibnamefont {Ömer Civalek}}\ and\
  \bibinfo {author} {\bibnamefont {Çiĝdem Demir}},\ }\bibfield  {title}
  {\bibinfo {title} {Bending analysis of microtubules using nonlocal
  euler-bernoulli beam theory},\ }\bibfield  {journal} {\bibinfo  {journal}
  {Applied Mathematical Modelling}\ }\textbf {\bibinfo {volume} {35}},\ \href
  {https://doi.org/10.1016/j.apm.2010.11.004} {10.1016/j.apm.2010.11.004}
  (\bibinfo {year} {2011})\BibitemShut {NoStop}%
\bibitem [{\citenamefont {Orlandini}\ \emph {et~al.}(2013)\citenamefont
  {Orlandini}, \citenamefont {Marenduzzo},\ and\ \citenamefont
  {Goryachev}}]{Orlandini2013}%
  \BibitemOpen
  \bibfield  {author} {\bibinfo {author} {\bibfnamefont {E.}~\bibnamefont
  {Orlandini}}, \bibinfo {author} {\bibfnamefont {D.}~\bibnamefont
  {Marenduzzo}},\ and\ \bibinfo {author} {\bibfnamefont {A.~B.}\ \bibnamefont
  {Goryachev}},\ }\bibfield  {title} {\bibinfo {title} {Domain formation on
  curved membranes: Phase separation or turing patterns?},\ }\bibfield
  {journal} {\bibinfo  {journal} {Soft Matter}\ }\textbf {\bibinfo {volume}
  {9}},\ \href {https://doi.org/10.1039/c3sm50650a} {10.1039/c3sm50650a}
  (\bibinfo {year} {2013})\BibitemShut {NoStop}%
\bibitem [{\citenamefont {Okabe}\ \emph {et~al.}(2008)\citenamefont {Okabe},
  \citenamefont {Xu},\ and\ \citenamefont {Burdine}}]{Okabe2008}%
  \BibitemOpen
  \bibfield  {author} {\bibinfo {author} {\bibfnamefont {N.}~\bibnamefont
  {Okabe}}, \bibinfo {author} {\bibfnamefont {B.}~\bibnamefont {Xu}},\ and\
  \bibinfo {author} {\bibfnamefont {R.~D.}\ \bibnamefont {Burdine}},\
  }\bibfield  {title} {\bibinfo {title} {Fluid dynamics in zebrafish kupffer's
  vesicle},\ }\bibfield  {journal} {\bibinfo  {journal} {Developmental
  Dynamics}\ }\textbf {\bibinfo {volume} {237}},\ \href
  {https://doi.org/10.1002/dvdy.21730} {10.1002/dvdy.21730} (\bibinfo {year}
  {2008})\BibitemShut {NoStop}%
\bibitem [{\citenamefont {Gutzeit}\ and\ \citenamefont
  {Koppa}(1982)}]{Gutzeit1982}%
  \BibitemOpen
  \bibfield  {author} {\bibinfo {author} {\bibfnamefont {H.~O.}\ \bibnamefont
  {Gutzeit}}\ and\ \bibinfo {author} {\bibfnamefont {R.}~\bibnamefont
  {Koppa}},\ }\bibfield  {title} {\bibinfo {title} {Time-lapse film analysis of
  cytoplasmic streaming during late oogenesis of drosophila},\ }\bibfield
  {journal} {\bibinfo  {journal} {Journal of Embryology and Experimental
  Morphology}\ }\textbf {\bibinfo {volume} {Vol. 67}},\ \href
  {https://doi.org/10.1242/dev.67.1.101} {10.1242/dev.67.1.101} (\bibinfo
  {year} {1982})\BibitemShut {NoStop}%
\bibitem [{\citenamefont {Alsous}\ \emph {et~al.}(2021)\citenamefont {Alsous},
  \citenamefont {Romeo}, \citenamefont {Jackson}, \citenamefont {Mason},
  \citenamefont {Dunkel},\ and\ \citenamefont {Martin}}]{Alsous2021}%
  \BibitemOpen
  \bibfield  {author} {\bibinfo {author} {\bibfnamefont {J.~I.}\ \bibnamefont
  {Alsous}}, \bibinfo {author} {\bibfnamefont {N.}~\bibnamefont {Romeo}},
  \bibinfo {author} {\bibfnamefont {J.~A.}\ \bibnamefont {Jackson}}, \bibinfo
  {author} {\bibfnamefont {F.~M.}\ \bibnamefont {Mason}}, \bibinfo {author}
  {\bibfnamefont {J.}~\bibnamefont {Dunkel}},\ and\ \bibinfo {author}
  {\bibfnamefont {A.~C.}\ \bibnamefont {Martin}},\ }\bibfield  {title}
  {\bibinfo {title} {Dynamics of hydraulic and contractile wave-mediated fluid
  transport during drosophila oogenesis},\ }\bibfield  {journal} {\bibinfo
  {journal} {Proceedings of the National Academy of Sciences of the United
  States of America}\ }\textbf {\bibinfo {volume} {118}},\ \href
  {https://doi.org/10.1073/pnas.2019749118} {10.1073/pnas.2019749118} (\bibinfo
  {year} {2021})\BibitemShut {NoStop}%
\bibitem [{\citenamefont {Lepesant}\ \emph {et~al.}(2024)\citenamefont
  {Lepesant}, \citenamefont {Roland-Gosselin}, \citenamefont {Guillemet},
  \citenamefont {Bernard},\ and\ \citenamefont {Guichet}}]{Lepesant2024}%
  \BibitemOpen
  \bibfield  {author} {\bibinfo {author} {\bibfnamefont {J.~A.}\ \bibnamefont
  {Lepesant}}, \bibinfo {author} {\bibfnamefont {F.}~\bibnamefont
  {Roland-Gosselin}}, \bibinfo {author} {\bibfnamefont {C.}~\bibnamefont
  {Guillemet}}, \bibinfo {author} {\bibfnamefont {F.}~\bibnamefont {Bernard}},\
  and\ \bibinfo {author} {\bibfnamefont {A.}~\bibnamefont {Guichet}},\
  }\bibfield  {title} {\bibinfo {title} {The importance of the position of the
  nucleus in \textit{Drosophila} oocyte development},\ }\href
  {https://doi.org/10.3390/cells13020201} {\bibfield  {journal} {\bibinfo
  {journal} {Cells}\ }\textbf {\bibinfo {volume} {13}},\ \bibinfo {pages} {201}
  (\bibinfo {year} {2024})}\BibitemShut {NoStop}%
\bibitem [{\citenamefont {Chakrabarti}\ \emph {et~al.}(2024)\citenamefont
  {Chakrabarti}, \citenamefont {Rachh}, \citenamefont {Shvartsman},\ and\
  \citenamefont {Shelley}}]{chakrabarti2024cytoplasmic}%
  \BibitemOpen
  \bibfield  {author} {\bibinfo {author} {\bibfnamefont {B.}~\bibnamefont
  {Chakrabarti}}, \bibinfo {author} {\bibfnamefont {M.}~\bibnamefont {Rachh}},
  \bibinfo {author} {\bibfnamefont {S.~Y.}\ \bibnamefont {Shvartsman}},\ and\
  \bibinfo {author} {\bibfnamefont {M.~J.}\ \bibnamefont {Shelley}},\
  }\bibfield  {title} {\bibinfo {title} {Cytoplasmic stirring by active
  carpets},\ }\href@noop {} {\bibfield  {journal} {\bibinfo  {journal}
  {Proceedings of the National Academy of Sciences}\ }\textbf {\bibinfo
  {volume} {121}},\ \bibinfo {pages} {e2405114121} (\bibinfo {year}
  {2024})}\BibitemShut {NoStop}%
\bibitem [{\citenamefont {Tracey}\ \emph {et~al.}(2000)\citenamefont {Tracey},
  \citenamefont {Ning}, \citenamefont {Klingler}, \citenamefont {Kramer},\ and\
  \citenamefont {Gergen}}]{tracey2000quantitative}%
  \BibitemOpen
  \bibfield  {author} {\bibinfo {author} {\bibfnamefont {J.}~\bibnamefont
  {Tracey}, \bibfnamefont {W.~D.}}, \bibinfo {author} {\bibfnamefont
  {X.}~\bibnamefont {Ning}}, \bibinfo {author} {\bibfnamefont {M.}~\bibnamefont
  {Klingler}}, \bibinfo {author} {\bibfnamefont {S.~G.}\ \bibnamefont
  {Kramer}},\ and\ \bibinfo {author} {\bibfnamefont {J.~P.}\ \bibnamefont
  {Gergen}},\ }\bibfield  {title} {\bibinfo {title} {Quantitative analysis of
  gene function in the \textit{Drosophila} embryo},\ }\href
  {https://doi.org/10.1093/genetics/154.1.273} {\bibfield  {journal} {\bibinfo
  {journal} {Genetics}\ }\textbf {\bibinfo {volume} {154}},\ \bibinfo {pages}
  {273} (\bibinfo {year} {2000})}\BibitemShut {NoStop}%
\bibitem [{\citenamefont {Prasad}\ \emph {et~al.}(2007)\citenamefont {Prasad},
  \citenamefont {Jang}, \citenamefont {Starz-Gaiano}, \citenamefont {Melani},\
  and\ \citenamefont {Montell}}]{prasad2007protocol}%
  \BibitemOpen
  \bibfield  {author} {\bibinfo {author} {\bibfnamefont {M.}~\bibnamefont
  {Prasad}}, \bibinfo {author} {\bibfnamefont {A.~C.}\ \bibnamefont {Jang}},
  \bibinfo {author} {\bibfnamefont {M.}~\bibnamefont {Starz-Gaiano}}, \bibinfo
  {author} {\bibfnamefont {M.}~\bibnamefont {Melani}},\ and\ \bibinfo {author}
  {\bibfnamefont {D.~J.}\ \bibnamefont {Montell}},\ }\bibfield  {title}
  {\bibinfo {title} {A protocol for culturing \textit{Drosophila melanogaster}
  stage 9 egg chambers for live imaging},\ }\href
  {https://doi.org/10.1038/nprot.2007.363} {\bibfield  {journal} {\bibinfo
  {journal} {Nature Protocols}\ }\textbf {\bibinfo {volume} {2}},\ \bibinfo
  {pages} {2467} (\bibinfo {year} {2007})}\BibitemShut {NoStop}%
\end{thebibliography}%


\begin{thebibliography}{8}%
\makeatletter
\providecommand \@ifxundefined [1]{%
 \@ifx{#1\undefined}
}%
\providecommand \@ifnum [1]{%
 \ifnum #1\expandafter \@firstoftwo
 \else \expandafter \@secondoftwo
 \fi
}%
\providecommand \@ifx [1]{%
 \ifx #1\expandafter \@firstoftwo
 \else \expandafter \@secondoftwo
 \fi
}%
\providecommand \natexlab [1]{#1}%
\providecommand \enquote  [1]{``#1''}%
\providecommand \bibnamefont  [1]{#1}%
\providecommand \bibfnamefont [1]{#1}%
\providecommand \citenamefont [1]{#1}%
\providecommand \href@noop [0]{\@secondoftwo}%
\providecommand \href [0]{\begingroup \@sanitize@url \@href}%
\providecommand \@href[1]{\@@startlink{#1}\@@href}%
\providecommand \@@href[1]{\endgroup#1\@@endlink}%
\providecommand \@sanitize@url [0]{\catcode `\\12\catcode `\$12\catcode
  `\&12\catcode `\#12\catcode `\^12\catcode `\_12\catcode `\%12\relax}%
\providecommand \@@startlink[1]{}%
\providecommand \@@endlink[0]{}%
\providecommand \url  [0]{\begingroup\@sanitize@url \@url }%
\providecommand \@url [1]{\endgroup\@href {#1}{\urlprefix }}%
\providecommand \urlprefix  [0]{URL }%
\providecommand \Eprint [0]{\href }%
\providecommand \doibase [0]{https://doi.org/}%
\providecommand \selectlanguage [0]{\@gobble}%
\providecommand \bibinfo  [0]{\@secondoftwo}%
\providecommand \bibfield  [0]{\@secondoftwo}%
\providecommand \translation [1]{[#1]}%
\providecommand \BibitemOpen [0]{}%
\providecommand \bibitemStop [0]{}%
\providecommand \bibitemNoStop [0]{.\EOS\space}%
\providecommand \EOS [0]{\spacefactor3000\relax}%
\providecommand \BibitemShut  [1]{\csname bibitem#1\endcsname}%
\let\auto@bib@innerbib\@empty
%</preamble>
\bibitem [{\citenamefont {Institute}(2023)}]{skellysim}%
  \BibitemOpen
  \bibfield  {author} {\bibinfo {author} {\bibfnamefont {F.}~\bibnamefont
  {Institute}},\ }\href {https://github.com/flatironinstitute/SkellySim}
  {\bibinfo {title} {Skellysim}} (\bibinfo {year} {2023}),\ \bibinfo {note}
  {accessed: 2023-12-25}\BibitemShut {NoStop}%
\bibitem [{\citenamefont {Dutta}\ \emph {et~al.}(2024)\citenamefont {Dutta},
  \citenamefont {Farhadifar}, \citenamefont {Lu} \emph {et~al.}}]{Dutta2024}%
  \BibitemOpen
  \bibfield  {author} {\bibinfo {author} {\bibfnamefont {S.}~\bibnamefont
  {Dutta}}, \bibinfo {author} {\bibfnamefont {R.}~\bibnamefont {Farhadifar}},
  \bibinfo {author} {\bibfnamefont {W.}~\bibnamefont {Lu}}, \emph {et~al.},\
  }\bibfield  {title} {\bibinfo {title} {Self-organized intracellular
  twisters},\ }\href {https://doi.org/10.1038/s41567-023-02372-1} {\bibfield
  {journal} {\bibinfo  {journal} {Nat. Phys.}\ }\textbf {\bibinfo {volume}
  {20}},\ \bibinfo {pages} {666} (\bibinfo {year} {2024})}\BibitemShut
  {NoStop}%
\bibitem [{\citenamefont {Fortunato}(2024)}]{fortunato2024high}%
  \BibitemOpen
  \bibfield  {author} {\bibinfo {author} {\bibfnamefont {D.}~\bibnamefont
  {Fortunato}},\ }\bibfield  {title} {\bibinfo {title} {A high-order fast
  direct solver for surface pdes},\ }\href@noop {} {\bibfield  {journal}
  {\bibinfo  {journal} {SIAM Journal on Scientific Computing}\ }\textbf
  {\bibinfo {volume} {46}},\ \bibinfo {pages} {A2582} (\bibinfo {year}
  {2024})}\BibitemShut {NoStop}%
\bibitem [{\citenamefont {Chakrabarti}\ \emph {et~al.}(2024)\citenamefont
  {Chakrabarti}, \citenamefont {Rachh}, \citenamefont {Shvartsman},\ and\
  \citenamefont {Shelley}}]{chakrabarti2024cytoplasmic}%
  \BibitemOpen
  \bibfield  {author} {\bibinfo {author} {\bibfnamefont {B.}~\bibnamefont
  {Chakrabarti}}, \bibinfo {author} {\bibfnamefont {M.}~\bibnamefont {Rachh}},
  \bibinfo {author} {\bibfnamefont {S.~Y.}\ \bibnamefont {Shvartsman}},\ and\
  \bibinfo {author} {\bibfnamefont {M.~J.}\ \bibnamefont {Shelley}},\
  }\bibfield  {title} {\bibinfo {title} {Cytoplasmic stirring by active
  carpets},\ }\href@noop {} {\bibfield  {journal} {\bibinfo  {journal}
  {Proceedings of the National Academy of Sciences}\ }\textbf {\bibinfo
  {volume} {121}},\ \bibinfo {pages} {e2405114121} (\bibinfo {year}
  {2024})}\BibitemShut {NoStop}%
\bibitem [{\citenamefont {Pozrikidis}(1992)}]{pozrikidis1992boundary}%
  \BibitemOpen
  \bibfield  {author} {\bibinfo {author} {\bibfnamefont {C.}~\bibnamefont
  {Pozrikidis}},\ }\href@noop {} {\emph {\bibinfo {title} {Boundary integral
  and singularity methods for linearized viscous flow}}}\ (\bibinfo
  {publisher} {Cambridge university press},\ \bibinfo {year}
  {1992})\BibitemShut {NoStop}%
\bibitem [{\citenamefont {Tanzosh}\ \emph {et~al.}(1992)\citenamefont
  {Tanzosh}, \citenamefont {Manga},\ and\ \citenamefont
  {Stone}}]{tanzosh1992boundary}%
  \BibitemOpen
  \bibfield  {author} {\bibinfo {author} {\bibfnamefont {J.}~\bibnamefont
  {Tanzosh}}, \bibinfo {author} {\bibfnamefont {M.}~\bibnamefont {Manga}},\
  and\ \bibinfo {author} {\bibfnamefont {H.~A.}\ \bibnamefont {Stone}},\
  }\bibfield  {title} {\bibinfo {title} {Boundary integral methods for viscous
  free-boundary problems: Deformation of single and multiple fluid-fluid
  interfaces},\ }\href@noop {} {\bibfield  {journal} {\bibinfo  {journal}
  {Boundary Element Technology VII}\ ,\ \bibinfo {pages} {19}} (\bibinfo {year}
  {1992})}\BibitemShut {NoStop}%
\bibitem [{\citenamefont {Rachh}\ \emph {et~al.}(2017)\citenamefont {Rachh},
  \citenamefont {Kl{\"o}ckner},\ and\ \citenamefont {O'Neil}}]{rachh2017fast}%
  \BibitemOpen
  \bibfield  {author} {\bibinfo {author} {\bibfnamefont {M.}~\bibnamefont
  {Rachh}}, \bibinfo {author} {\bibfnamefont {A.}~\bibnamefont
  {Kl{\"o}ckner}},\ and\ \bibinfo {author} {\bibfnamefont {M.}~\bibnamefont
  {O'Neil}},\ }\bibfield  {title} {\bibinfo {title} {Fast algorithms for
  quadrature by expansion i: Globally valid expansions},\ }\href@noop {}
  {\bibfield  {journal} {\bibinfo  {journal} {Journal of Computational
  Physics}\ }\textbf {\bibinfo {volume} {345}},\ \bibinfo {pages} {706}
  (\bibinfo {year} {2017})}\BibitemShut {NoStop}%
\bibitem [{\citenamefont {Corona}\ \emph {et~al.}(2017)\citenamefont {Corona},
  \citenamefont {Greengard}, \citenamefont {Rachh},\ and\ \citenamefont
  {Veerapaneni}}]{corona2017integral}%
  \BibitemOpen
  \bibfield  {author} {\bibinfo {author} {\bibfnamefont {E.}~\bibnamefont
  {Corona}}, \bibinfo {author} {\bibfnamefont {L.}~\bibnamefont {Greengard}},
  \bibinfo {author} {\bibfnamefont {M.}~\bibnamefont {Rachh}},\ and\ \bibinfo
  {author} {\bibfnamefont {S.}~\bibnamefont {Veerapaneni}},\ }\bibfield
  {title} {\bibinfo {title} {An integral equation formulation for rigid bodies
  in stokes flow in three dimensions},\ }\href@noop {} {\bibfield  {journal}
  {\bibinfo  {journal} {Journal of Computational Physics}\ }\textbf {\bibinfo
  {volume} {332}},\ \bibinfo {pages} {504} (\bibinfo {year}
  {2017})}\BibitemShut {NoStop}%
\end{thebibliography}%
\end{document}

\preprint{APS/123-QED}
\title{Supplemental Material: Geometric Effects in Large Scale Intracellular Flows}

\author{Olenka Jain}
\affiliation{Lewis- Sigler  Institute  for  Integrative  Genomics,  Princeton  University, Princeton, NJ 08544}
\affiliation{Center for Computational Biology, Flatiron Institute, New York, 10010, USA}
\author{Brato Chakrabarti}
\affiliation{International Center for Theoretical Sciences, Bengaluru, 560089, India}%
\author{Reza Farhadifar}
\affiliation{Center for Computational Biology, Flatiron Institute, New York, 10010, USA}
\author{Elizabeth R. Gavis}
\affiliation{Department of Molecular Biology, Princeton University, Princeton, NJ 08544}
\author{Michael J. Shelley}
\affiliation{Center for Computational Biology, Flatiron Institute, New York, 10010, USA}
\affiliation{Courant Institute, New York University, New York, NY 10012, USA}
\author{Stanislav Y. Shvartsman}
\affiliation{Center for Computational Biology, Flatiron Institute, New York, 10010, USA}
\affiliation{Department of Molecular Biology, Princeton University, Princeton, NJ 08544}
\affiliation{Lewis- Sigler  Institute  for  Integrative  Genomics,  Princeton  University, Princeton, NJ 08544}

\maketitle

\section*{Simulation of microtubules in ellipsoidal geometries}
We report the main governing equations for the discrete microtubule (MT) simulations from the main text:
\begin{align}
\eta (\mathbf{X}^i_{t}-\mathbf{u}) &= (\mathbf{I}+ \mathbf{X}_{s}^{i} \mathbf{X}_{s}^{i}) \cdot (\mathbf{f}^i - \sigma \mathbf{X}_{s}^{i}), 
\\
\mathbf{f}^i &= -E \mathbf{X}_{ssss}^{i} + (T^{i} \mathbf{X}_{s}^{i})_{s},
\end{align}
where $s$ is the arc-length of MT centerline, $\eta={8 \pi \mu}/{|c|}$, where $\mu$ is the viscosity, and $c=\log e \epsilon^2$ is a coefficient characterizing the slenderness of the MT ($\epsilon$ is the ratio of MT length to width). The flow of the cytoplasm is governed by the forced incompressible Stokes equation and subject to a no-slip boundary condition at the cortex: 
\begin{align}
\nabla q - \mu \Delta \mathbf{u} &=  \sum\limits_{i=1}^N \int\limits_{0}^{L} \mathrm{d}s \  \mathbf{f}^{i}(\textit{s})\delta(\mathbf{x}-\mathbf{X}^{i}), \\
\nabla \cdot \bu &= 0,
\end{align}
where $q$ is pressure and $\mathbf{u}(\bx)$ is the cytoplasmic velocity field. We solve these coupled fiber and fluid equations using the open-source computational platform SkellySim \cite{skellysim,Dutta2024}.

For the simulations in spheres and ellipsoids, the anchoring points of the MTs were uniformly distributed using: (a) a Fibonacci lattice method and (b) randomly with the constraint that no two microtubules were within at least two MT lengths of each other. The microtubules were initialized normally to the surface of the ellipsoid. Ellipsoidal peripheries of aspect ratios from $1.2-2.2$ were generated, matching the measured aspect ratios of stage 12 oocytes. The simulations were evolved until they reached a plateau in the total elastic energy, which was considered an indication that the system had evolved to a steady state. To quantify the elastic energy, the curvature $\kappa_{i}(s,t)$ of each fiber was calculated using a $2$nd order finite difference scheme. The net elastic energy was then calculated as
\begin{eqnarray}
\mathcal{E}(t) =  \sum_{i=1}^{N} \frac{E}{2} \int_{0}^{L}  ds[\kappa_i(s,t)]^2.
\end{eqnarray}

To quantify defects on the MT bed we used the polarity field:
\begin{equation}
\mathbf{p}^i (t) = (\mathbf{I} - \mathbf{nn}) \cdot \frac{\mathbf{X}^i (L,t) - \mathbf{X}^i (0,t)}{L},
\end{equation}
where $\bn$ is the inward unit surface normal at $\bX^{i}(0,t)$. We interpolated these polarity vectors onto a structured grid on the surface. The minimum values of the resulting interpolated polarity field provided us with the location of the two defects. Additionally, we used the open-source software SurfaceFun \cite{fortunato2024high} to perform a Hodge decomposition of this interpolated surface vector field. The defects were identified as the maxima in the absolute value of the divergence-free component of the Hodge decomposition. Both of these methods produced the same result. %Flow orientation was characterized by the position of the defects in the microtubule order for two reasons. Firstly, because the flow is Stokes and driven entirely by the motion of the MTs and motors, the entire flow field could be represented by the instantaneous position of the MTs. Secondly, because the resulting flow field was largely solid body rotation and the axis of rotation was defined by the line connecting the two defects, the defect positions determined the axis of rotation of flow.  

For our simulated flow perturbations, the computations were first evolved in a sphere until a steady state was determined by a plateau in elastic energy.  Using the emergent steady state as our initial conditions, we changed the geometry of the sphere to an ellipsoid with the same surface area and an axis transverse to the defect axis. Each fiber was rotated such that the base was clamped normally to the cortex to satisfy the boundary conditions.

\section*{Active Carpet Model}

The details of the formulation and governing equations for the active carpet model have been outlined in \cite{chakrabarti2024cytoplasmic}. Here, we outline how the cytoplasmic fluid velocity $\bu(\bx)$ is computed in this model in the presence of a nucleus that is modeled as a stationary no-slip surface. The fluid flow is governed by the following equations 
\begin{alignat}{2}
    -\nabla q + \Delta \bu = \mathbf{0}, &  \ \ \nabla \cdot \bu  = 0, \label{eq:homst}& \\ 
    \bu(\by,t) = \mathbf{0}, &  \ \ \text{ on } \partial D \  \text{and}  \ \partial N \\ 
    \llbracket \bu \rrbracket = \mathbf{0}, \ \ \llbracket \boldsymbol{\sigma}  \cdot \bn \rrbracket (\by,t) = -\bar{\bff}(\by,t).& \ \ \text{ on } \partial S. & \label{eq:stokeappx}
\end{alignat}
Here $\partial D$ is the surface of the cell cortex, $\partial N$ is the surface of the nucleus, and $\partial S = \partial D + \bn$ is the surface across which the stress jump $\bar{\bff}$ acts. The solution to the above equations are recast as an integral equation of the form \cite{pozrikidis1992boundary,tanzosh1992boundary}:
\begin{equation}\label{eq:intfor}
        \bu(\bx) = \frac{1}{8 \pi} \int_{\partial S} \bar{\bff}(\by) \cdot \mathcal{G}(\bx,\by) \ \md A(\by) -\frac{1}{8 \pi} \int_{\partial D} \bff^w(\by) \cdot \mathcal{G}(\bx,\by) \ \md A(\by) - \frac{1}{8 \pi} \int_{\partial N} \bff^n(\by) \cdot \mathcal{G}(\bx,\by) \ \md A(\by),
\end{equation}
where $\mathcal{G}(\bx,\by)$ is the free-space Green's function for the Stokes equation. In Eq.~\eqref{eq:intfor},
$\bff^w(\by)$ and $\bff^n(\by)$  are unknown tractions on the cell cortex $\partial D$ and $\partial N$. Both of these tractions are determined by imposing the no-slip condition $\bu(\by) = \mathbf{0}$ on $\partial D$ and $\partial N$. This boundary-integral formulation allows fast simulations of flows inside smooth closed surfaces. We use a 5th-order accurate quadrature scheme to discretize the surface integrals of Eq.~\eqref{eq:intfor} and use quadrature by expansion to evaluate near interactions as outlined in \cite{rachh2017fast,corona2017integral}. For simulations presented here, we used $M_e \sim 2700$ quadrature points on the ellipsoidal surface and $M_n \sim 900$ points on the surface of the nucleus.

%----------------------------------------------------------------------------------------
%	BIBLIOGRAPHY
%----------------------------------------------------------------------------------------

%\newpage

%\bibliography{GeomControl_1} % Use the NIHGrant.bib file for the reference list, replace with your own

%\bibliographystyle{nihunsrt} % Use the custom nihunsrt bibliography style included with the template

%----------------------------------------------------------------------------------------
\bibliography{bibfile}% Produces the bibliography via BibTeX.